\documentclass[twocolumn,superscriptaddress,nobibnotes,footinbib,amsmath,floatfix,aps,prapplied,citeautoscript,reprint]{revtex4-1}

\usepackage{graphicx}
\usepackage{dcolumn}
\usepackage{bm}
\usepackage{amsmath,bm,multirow}
\usepackage{amsfonts}
\usepackage{float}
\usepackage{color}
\usepackage[draft]{changes} 
\usepackage[normalem]{ulem}
\definechangesauthor[name={yi}, color=blue]{Yi}

\newcommand{\psu}{Department of Mechanical and Materials Engineering, Portland State University, Portland, OR 97201, USA}

\usepackage{float}
\usepackage{graphicx}
\usepackage{sidecap}
\usepackage{tabularx}
\usepackage{sidecap}

\begin{document}
\title{Machine Learning a Universal Harmonic Interatomic Potential for Predicting Phonons in Crystalline Solids}

\author{Huiju Lee}
\affiliation{\psu}

\author{Yi Xia}
\email{yxia@pdx.edu}
\affiliation{\psu}

\date{\today}

\begin{abstract}
Phonons, as quantized vibrational modes in crystalline materials, play a crucial role in determining a wide range of physical properties, such as thermal and electrical conductivity, making their study a cornerstone in materials science. 
In this study, we present a simple yet effective strategy for deep learning harmonic phonons in crystalline solids by leveraging existing phonon databases and state-of-the-art machine learning techniques.
The key of our method lies in transforming existing phonon datasets, primarily represented in interatomic force constants, into a force-displacement representation suitable for training machine learning universal interatomic potentials.
By applying our approach to one of the largest phonon databases publicly available, we demonstrate that the resultant machine learning universal harmonic interatomic potential not only accurately predicts full harmonic phonon spectra but also calculates key thermodynamic properties with remarkable precision. 
Furthermore, the restriction to a harmonic potential energy surface in our model provides a way of assessing uncertainty in machine learning predictions of vibrational properties, essential for guiding further improvements and applications in materials science.

\end{abstract}
\maketitle

\begin{figure*}[htp]
	\includegraphics[width = 0.8\linewidth]{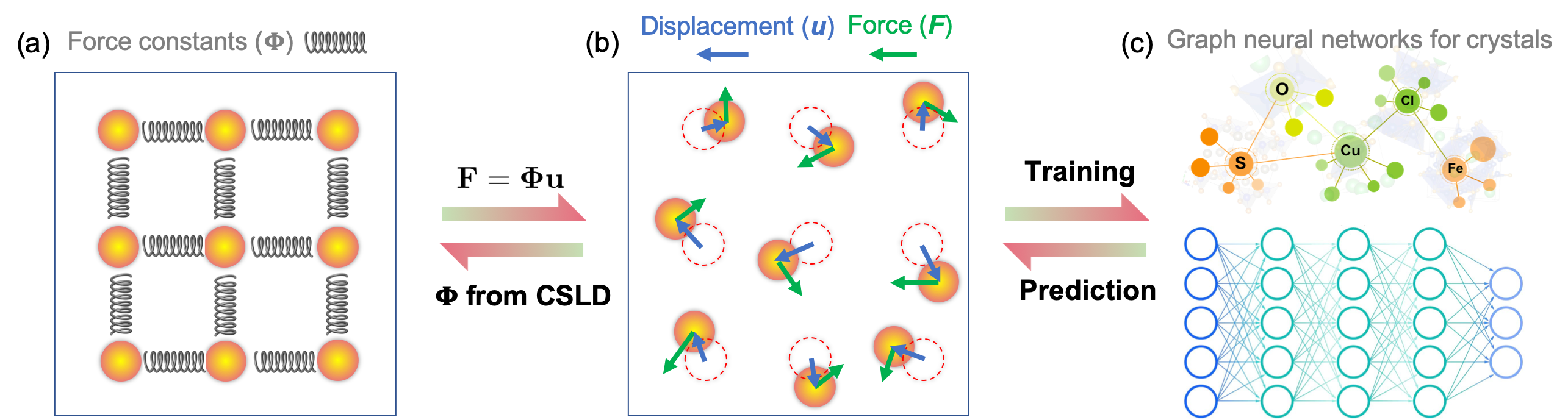}
	\caption{
	A schematic showing the conversions between different representations for describing interatomic interactions. (a) Interatomic force constants (IFCs) representation, wherein atoms and IFCs are indicated by solid spheres and springs, respectively; (b) force-displacement (FD) representation, wherein dashed circles and solid spheres are used to denote equilibrium and displaced atom positions, respectively; (c) Graph neural networks (GNNs) representation, wherein crystal structures are represented by graphs with atoms as nodes and bonds between atoms as edges.
	}
	\label{fig:flowchat}
\end{figure*}


All materials are made up of atoms, which vibrate ubiquitously, even at absolute zero temperature, due to quantum effects. When present in periodic solids, these vibrations can be quantized as quasiparticles, known as phonons. The study of phonons is an integral part of solid-state physics and materials science, as they play an essential role in many physical properties of materials, including thermodynamic stability, thermal conductivity, and electric conductivity~\cite{wallace1998thermodynamics}. Specifically, concerning the fundamental thermodynamic properties, phonons are vital for materials' finite-temperature characteristics, such as thermal expansion, heat capacity, free energy, and lattice stability. Phonons also serve as a gene governing energy transfer in solids, manifested in lattice heat transfer~\cite{wallace1998thermodynamics,QIAN2021154}. Furthermore, when coupled with other quasiparticles, e.g., electrons, phonons can strongly modulate carrier conductivity and give rise to conventional superconductivity~\cite{RevModPhys.89.015003}. However, the properties of the materials currently contained in existing databases~\cite{matproj,aflow,oqmd} are limited to those obtained by relatively simple ground state calculations – formation energies, electronic band-gaps, and -structures, etc. – with no dynamical information such as phonons. The most extensive collection to date of vibrational properties contains only several thousands of compounds~\cite{Petretto2018,phdb2018}, which are focused on the harmonic vibrational properties. This poses a crucial limitation regarding the prediction of materials at ambient or higher temperatures.

Data-driven, or machine learning (ML), approaches are becoming a method of choice for materials design, discovery, and property prediction, thanks to their extraordinary capability of modeling complex composition-structure-property mapping in materials~\cite{ramprasad2017machine,griesemer2023accelerating}. Consequently, efforts are now being devoted to developing ML models encoded with vibrational properties to overcome the challenge due to the scarcity of data. The promising applications of ML approaches to modeling phonons fall into two categories. The first category highlights ML models that use either hand-crafted descriptors or learned representations to directly predict phonon properties, such as vibrational frequency, density-of-states (DOS), and vibrational free energy, or incorporate such information implicitly when predicting materials stability without resorting to constructing interatomic potential~\cite{bartel2018physical,zhantao,Nguyen2022,Gurunathan}. For example. Chen \textit{et al.}~\cite{zhantao} employed the Euclidean neural network to directly predict phonon DOS using a  training set of about 1000 examples with over 64 atom types. By capturing full crystal symmetry, their model reproduces key features of experimental data and even generalizes to materials with unseen elements. Similar results are demonstrated by Gurunathan \textit{et al.}~\cite{Gurunathan} using the atomistic line graph neural network (ALIGNN) on a significantly expanded database composed of over 14000 phonon spectra computed at the Brillouin center. In another study, Nguyen \textit{et al.}~\cite{Nguyen2022} used deep graph neural networks to predict lattice vibrational frequencies. Despite displaying low transferability across different structure types, their model implies the capability of deep graph neural networks to learn to predict lattice vibrational frequency when sufficient training samples are available.

Another category of ML approaches to modeling phonons relies on directly constructing machine learning interatomic potential (MLIP). The key idea is to learn the statistical relation between structure and potential energy without knowledge about the relevant interactions. Recent advances lead to an extensive collection of MLIPs, including kernel-based learning approaches and artificial neural networks, such as Behler-Parrinello neural network potentials (NNP)~\cite{PhysRevLett.98.146401}, Gaussian approximation potentials (GAP)~\cite{bartok2010gaussian}, and moment tensor potentials (MTP)~\cite{shapeev2016moment}. Most recently, deep learning representations using graph neural networks (GNNs) have been leveraged to develop machine learning universal interatomic potentials (MLUIPs) for arbitrary chemical species and structures. Specific realizations include but are not limited to, the crystal graph convolutional neural networks (CGCNN)~\cite{cgcnn}, the graph convolution network with continuous-filter convolutional layers  (SchNet)~\cite{schnet}, the directional message passing neural network (DimeNet)~\cite{dimenet}, and the geometric message passing neural network (GemNet)~\cite{gemnet}. Particularly, the material graph with three-body interactions neural network (M3GNet)~\cite{Chen2022} model, which has been trained on density functional theory (DFT)~\cite{dft1,dft2} crystal structure relaxations from the Materials Project~\cite{matproj}, demonstrates the potential to directly predict harmonic phonon dispersions. However, the accuracy of the M3GNet model is compromised despite covering broad chemistry, showing relatively large deviations of about one Thz for averaged phonon frequencies when compared to DFT~\cite{Chen2022}.


Applying these MLUIPs to directly modeling phonons is appealing. Meanwhile, challenges arise when one aims to train highly accurate MLUIPs due to the limitations of available training datasets. For example, the DFT crystal structure relaxations used to train the M3GNet model consist of only small unit/primitive cells and suffer from relatively low DFT convergence criteria, typically used for constructing large materials databases. Therefore, the resultant MLUIPs are intrinsically unreliable in yielding longer-range interatomic interactions and accurate forces, which are crucial to converge phonon calculations. To overcome these limitations, in this letter, we propose a strategy to leverage existing phonon databases and convert these databases into a proper form suitable for training MLUIPs. This is inspired by phonon databases usually being constructed from high-quality DFT calculations of supercell structures, which can better capture the longer-range interatomic interactions and exhibit accurate forces. Afterward, we focus on constructing a specific form of MLUIPs: a machine learning universal harmonic interatomic potential (MLUHIP) to deep learning phonons. We show that such a restriction of the potential energy surface to the harmonic form not only enables efficient model parameterization but also provides physics-inspired uncertain quantification for phonon predictions.


The key of our proposed strategy relies on conversions among different representations for describing interatomic interactions. For example, as shown in Fig.~\ref{fig:flowchat}(a), existing phonon databases typically adopt an interatomic force constants (IFCs) representations. In contrast, a rather different representation, for example, graph neural networks (GNNs), is used by MLUIPs, as shown in Fig.~\ref{fig:flowchat}(c). To bridge the gap between the IFCs and GNNs representation, we propose to create an intermediate force-displacement (FD) representation, shown in Fig.~\ref{fig:flowchat}(b). The conversion between different representations can be achieved as follows: the IFCs representation can be converted into the FD representation by taking the tensor product of IFCs and random atomic displacements, while the back conversion can be achieved by fitting IFCs using FD dataset, for instance, via the Compressive Sensing Lattice Dynamics (CSLD) approach~\cite{csld,csld1}. Meanwhile, the conversion between the FD and the GNNs representations can be realized by training GNNs-based MLUIPs and conducting predictions. Note that when the IFCs are limited to the second order, as in this study, a harmonic version of GNNs-based MLUIP (i.e., MLUHIP) will be pursued.

\begin{figure*}[htp]
	\includegraphics[width = 1.0\linewidth]{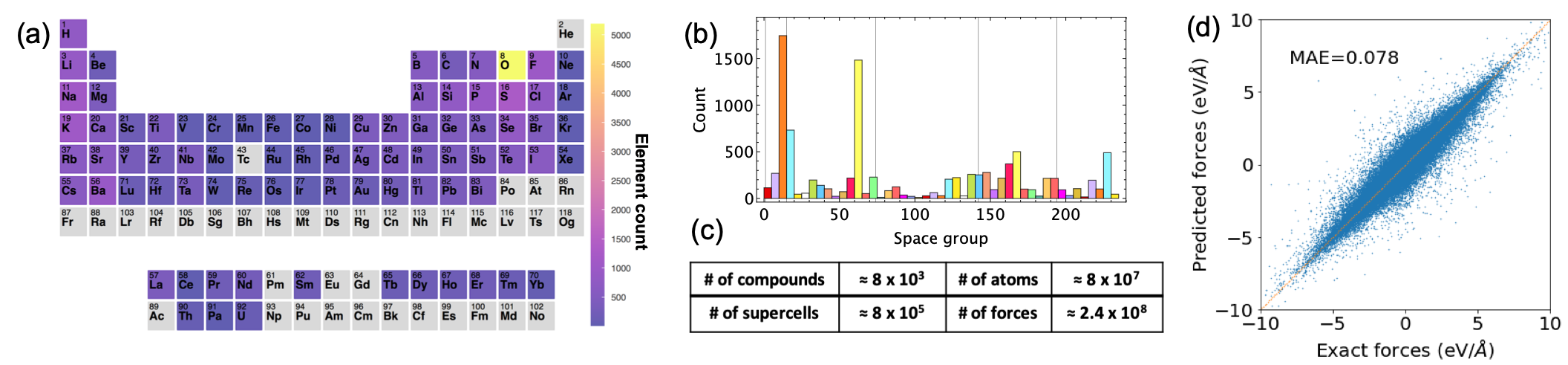}
	\caption{
	(a) A map of the frequencies of elements contained in the phonon database across the periodic table, covering 81 elements in total. (b) Distribution of symmetries (space group numbers) of the compounds in the database. (c) A table summarizing the information of the generated force-displacement dataset. (d) Predicted forces using our MLUHIP model compared with the exact harmonic forces computed using the harmonic interatomic force constants for the testing dataset.
	}
	\label{fig:datatrain}
\end{figure*}



To evaluate the feasibility of our strategy for deep learning harmonic phonons, we developed a GNNs-based MLUHIP to model harmonic phonons in crystalline solids. We used one of the largest phonon databases, i.e., the phonon database at Kyoto University~\cite{phdb2018}, to generate the FD training dataset for GNNs-based MLUHIP models. We chose this database because it contains DFT calculations of harmonic phonons for approximately 10000 inorganic solid compounds. These compounds cover a broad spectrum of chemistries and structures. Specifically, there are in total 81 chemical elements across the periodic table included in these compounds, and the resultant elemental distributions are detailed in Fig.~\ref{fig:datatrain}(a). The corresponding crystal structures also display a diverse distribution of symmetries, as indicated by the various space group numbers in Fig.~\ref{fig:datatrain}(b). It is worth noting that it is crucial to attain such a diverse dataset in order to develop an ML model that is potentially applicable to solids across the periodic table.

Next, we prepared a broad dataset for training a GNNs-based MLUHIP with the following procedures. First, we carefully screened all and down-selected compounds whose squared phonon frequencies are all non-negative to ensure dynamical stability. This step ensures that the compounds entering the training dataset are all dynamically stable, thus making the harmonic phonon approximation valid. This screening leads to a total number of 8229 compounds out of the original dataset. Second, since the calculated harmonic phonon for each compound in the original dataset is represented using IFCs, we then converted the IFCs representation to the FD representation by randomly perturbing atoms around their equilibrium positions, as illustrated in Fig.~\ref{fig:flowchat}(b). Specifically, we used the quantum covariance matrix to generate random atomic displacements at 300~K following the phonon dispersion~\cite{SSCHA,pbte2018} (see supplemental material). This approach is superior to an alternative method based on small displacements ($\approx$ 0.01 \r{A}). It is because the latter fails to account for temperature- and atom-dependent displacements. In order to obtain a comprehensive force-displacement dataset, we generate random atomic displacements for each compound over 100 supercell configurations, whose sizes are consistent with the supercell used for DFT phonon calculations. This results in an enormous dataset consisting of about 80 million atoms and 240 million force components, as summarized in Fig.~\ref{fig:datatrain}(c).  Finally, we randomly split the dataset into a training dataset, which contains 7829 compounds, out of which 411 compounds are selected for validation, and an additional testing dataset consisting of 400 compounds. Note that we split the force-displacement dataset by compounds instead of supercell configurations so there is no overlap between the compounds used for training and testing, thus ensuring more stringent criteria for validating the resultant ML models.


To train a GNNs-based MLUHIP using the constructed FD dataset, we employed the Directional Message Passing Neural Network (DimeNet) model. DimeNet was initially introduced by Gasteiger et al.~\cite{dimenet} to embed directional information between different atoms in a molecular or crystal during the message passing. This strategy has been shown to significantly improve the accuracy of predictions for molecular properties and use in molecular dynamics simulations. We used an improved version of DimeNet, i.e., DimeNet++~\cite{dimenet++}, for the model training to balance accuracy and efficiency. Moreover, we took advantage of an existing implementation of DimeNet++ in the Open Catalyst Project (OCP)~\cite{ocp_dataset}, which implements various kinds of MLUIP algorithms for different tasks and provides optimized parameters for training MLUIP models depending on the size of the dataset. Considering that our dataset contains more than 10 million atoms and our task is to predict forces given structures, we adopted a model specification for a complex dataset as provided by the OCP (see supplemental material). The model was then trained by minimizing the force prediction error on the training dataset for 15 epochs, a number that is consistent with tasks of similar size within OCP~\cite{ocp_dataset}. The trained model achieves a force prediction mean absolute error (MAE) of 0.049 eV/\r{A} on the validating dataset (see supplemental material) and 0.078 eV/\r{A} on the testing dataset, respectively; the latter is shown in Fig.~\ref{fig:datatrain}(d). Note that our testing error is similar to the value of 0.072 eV/\r{A} reported by the M3GNet model~\cite{Chen2022}. The relatively large error in force prediction may be attributed to several factors, including: (i) the use of a finite cutoff distance for interatomic interactions, which may omit relevant contributions from atoms just beyond this threshold; (ii) the omission of correlations beyond three-body interactions~\cite{dimenet++}, which can play a significant role in complex systems; and (iii) the disregard for potential anharmonic interactions that become significant with large atomic displacements. These aspects, among others, could contribute to discrepancies in predictive accuracy.



\begin{figure*}[htp]
	\includegraphics[width = 1.0\linewidth]{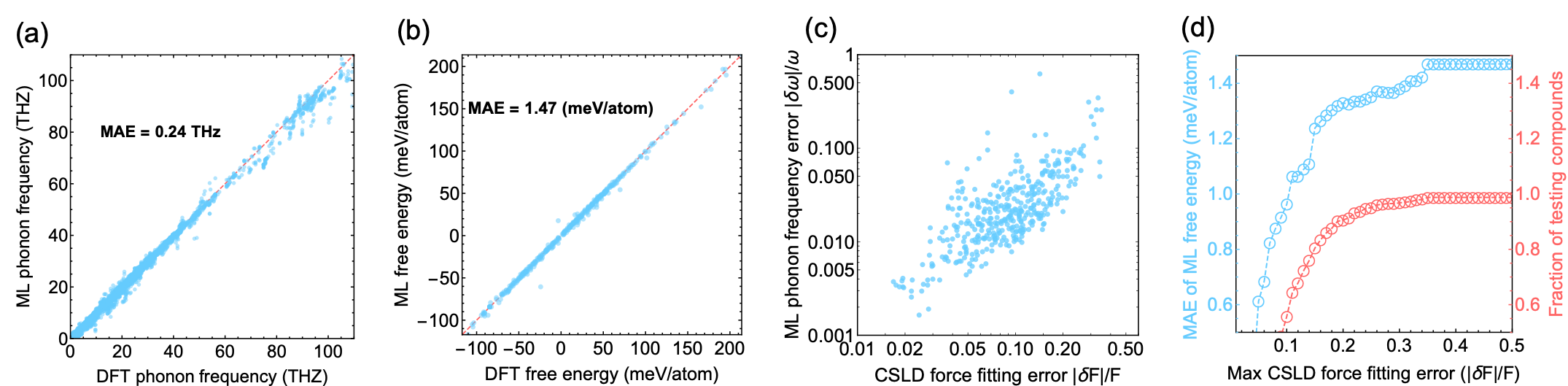}
	\caption{ Performance of the ML model on a testing dataset containing 400 unseen compounds.
	(a) Predicted phonon frequencies using our trained MLHUIP model compared with direct DFT calculations. (b) Predicted vibrational free energy using our trained MLHUIP model compared with the DFT results at 300 K. (c) Correlation between the harmonic force fitting error ($|\delta F|/F$) from CSLD using the forces predicted by the trained MLHUIP model and the relative error in predicted phonon frequency ($|\delta \omega|/\omega$). (d) Mean absolute error (MAE) of the machine learned free energy (left vertical axis in blue) for the compounds with a given maximum value of $|\delta F|/F$. Fraction of the total 400 compounds (right vertical axis in red) for a given maximum value of $|\delta F|/F$. 
	}
	\label{fig:dataML}
\end{figure*}

While our trained MLUHIP model predicts reasonably accurate forces for unseen compounds, its capability of predicting the full harmonic phonon spectrum needs to be assessed. To directly compare the predicted and the DFT-calculated phonons, we utilized CSLD~\cite{csld,csld1} to extract the IFCs using the forces predicted by our MLUHIP  model for the testing dataset. We chose not to use the finite displacement approach implemented in phonon software such as Phonopy~\cite{phonopy-phono3py-JPSJ} because it is more sensitive to force errors due to the adopted small atomic displacements. The resulting IFCs were then used to construct the dynamical matrix, diagonalized to obtain phonon frequencies, referred to as ML phonon frequency. As shown in Fig.~\ref{fig:dataML}(a), the ML phonon frequency compares well with the DFT phonon frequency, achieving an MAE of 0.24 THz (see supplemental material for the comparison of full phonon dispersions for the testing dataset). These results are encouraging considering the fact that the ML model has never seen the compounds in the testing dataset and meanwhile they exhibit broad distribution of chemistries and structures. Using the ML phonon frequency, we also computed various thermodynamic properties such as vibrational free energy, vibrational entropy, and heat capacity. The comparison between ML and DFT vibrational free energies at 300~K is shown in Fig.~\ref{fig:dataML}(b), showing an MAE of only about 1.45 meV/atom. Such an error is significantly smaller than previous models~\cite{Gurunathan,legrain2017chemical} and close to the criteria of approximately $\pm$1.0 meV/atom for predicting accurate phase transition temperatures~\cite{TUTILD}.

Having validated our MLUHIP model, an important question arises concerning the uncertainty estimation when applying the model to unseen datasets/materials. The capability of obtaining reliable uncertainty estimation is of great importance because it can be used to determine whether or not additional DFT calculations are required to improve model accuracy or perform active learning. We find that the restriction of potential energy surface to the harmonic order, as reflected in our MLUHIP model, naturally leads to an effective assessment of model uncertainty. This is because any additional terms beyond the quadratic form that appear in the model's predicted forces could be viewed as model uncertainties. Moreover, such uncertainties can be directly quantified by estimating the force prediction error when we fit the FD dataset by truncating at harmonic IFCs. For example, within the CSLD framework, the relative CSLD force fitting error ($ |\delta F|/F \equiv ||\mathbf{F}-\Phi^{\rm CS}\mathbf{u}||_2/ || \mathbf{F}||_2$, wherein $\mathbf{F}$ are predicted forces by an MLUHIP model, $\mathbf{u}$ are random atomic displacements, and $\Phi^{\rm CS}$ are fitted harmonic IFCs) can be used to indicate the model uncertainty, for instance, the error in predicted phonon frequency. Indeed, Fig.~\ref{fig:dataML}(c) shows that there is a strong correlation between $|\delta F|/F$ and the relative error in predicted phonon frequency ($|\delta \omega|/\omega$). Such a strong correlation can also be found between $|\delta F|/F$ and the MAE of the predicted vibrational free energy. Fig.~\ref{fig:dataML}(d) shows the cumulative distribution of errors for vibrational free energy as a function of maximum $|\delta F|/F$. It can be inferred that we can significantly reduce the MAE of predicted vibrational free energy by downselecting compounds with a given maximum value of $|\delta F|/F$. For instance, by limiting $|\delta F|/F$ to be about 0.1, corresponding to about half of the total calculated compounds, we achieve a reduced MAE error of only about 1.0 meV/atom for vibrational free energy.

Before closing, we briefly comment on future research directions. Building on the current research, potential work could explore several promising directions. First, the implementation of more advanced machine learning models, such as equivariant neural networks with inherent symmetry~\cite{thomas2018tensor,batzner20223,Batatia2022mace}, could potentially enhance the accuracy and generalizability of phonon predictions. Additionally, an important extension of this work could involve including anharmonic properties in our models to account for the significant impact of anharmonic effects on the thermal properties of materials, for example, via $\Delta$ machine learning~\cite{Ramakrishnan2015} on top of our harmonic model.
Applying our current machine learning model to a specific, targeted materials system and employing active learning techniques could further testify and refine its predictive capabilities. Moreover, the application of our approach to complex materials such as amorphous or disordered solids requires special attention due to the lack of periodicity in atomic arrangement and the presence of diverse local environments, a challenge recently has been recognized in predicting atomic-scale stiffness in an amorphous solid system~\cite{PENG2021101446}.
Another crucial area for future research lies in enhancing the efficiency of the CSLD approach. Improvements in CSLD could expedite the process of extracting interatomic force constants from ML-predicted forces, thereby streamlining the entire workflow from phonon database conversion to practical phonon spectrum prediction on a large scale.Finally, we envision that our model is particularly suited for calculating phonons in complex crystalline materials, where the computational demands are substantial. For small molecule materials with well-defined interatomic interactions, traditional methods like molecular dynamics (MD) may serve as a practical alternative. MD not only provides an efficient solution but also offers the advantage of incorporating anharmonic effects, which are essential for a comprehensive understanding of material behaviors.

In conclusion, our study demonstrates a simple yet effective strategy for deep learning harmonic phonons using machine learning universal interatomic potential based on graph neural networks and existing phonon databases. By converting the interatomic force constants representation from existing phonon databases into a more machine learning-friendly force-displacement representation, we could train our model on a large and diverse dataset, ensuring its applicability across various materials. The model's performance in predicting harmonic phonon frequencies and fundamental thermodynamic properties confirms its high accuracy and potential for widespread application in the field of materials science. Additionally, our approach to uncertainty quantification, inherent in our harmonic potential energy surface model, represents a significant step forward in the reliable application of machine learning models to materials beyond those included in the training dataset. This capability to estimate uncertainty is crucial for identifying the need for further data refinement or model improvement, paving the way for more informed and efficient research in materials design and discovery involving vibrational properties.

\section*{Supplementary Material}

See supplementary material for more detailed descriptions of the validation of the machine learning model, the compressive sensing lattice dynamics approach, the equation for generating temperature-dependent atomic displacements, and the comparison of phonon dispersions for the testing dataset.

\begin{acknowledgments}
H. L. and Y. X. acknowledge support from the US National Science Foundation through award 2317008. We acknowledge the computing resources provided by Bridges2 at Pittsburgh Supercomputing Center (PSC) through allocations mat220006p and mat220008p from the Advanced Cyber-infrastructure Coordination Ecosystem: Services \& Support (ACCESS) program, which is supported by National Science Foundation grants \#2138259, \#2138286, \#2138307, \#2137603, and \#2138296.
\end{acknowledgments}



\appendix

\bibliography{aipsamp}

\end{document}